\documentclass[intlimits,twoside,a4paper]{article}

\usepackage{amsmath,amssymb}
\usepackage{graphicx}

\usepackage[T2A]{fontenc}
\usepackage[cp1251]{inputenc}

\usepackage[eqsecnum]{cmpj2}
\issue{2016}{19}{4}{43001}
\doinumber{10.5488/CMP.19.43001}
\title[Effect of film thickness on the width of percolation threshold]%
{Effect of film thickness on the width of percolation threshold in metal-dielectric composites}
\vspace{2mm}

\author[M. Mokhtari, L. Zekri, A. Kaiss, N. Zekri]{M. Mokhtari\refaddr{label1,label2},
         L. Zekri\refaddr{label1}, A. Kaiss\refaddr{label3}, N. Zekri\refaddr{label1} }
\vspace{2mm}
\addresses{
\addr{label1} Universit\'e des Sciences et de la Technologie d'Oran Mohamed Boudiaf, USTO-MB, LEPM,\\ BP 1505, El M' Naouar, 31000 Oran, Algeria
\addr{label2} Centre Universitaire de  Tissemsilt, BP 182, Tissemsilt, Algeria
\addr{label3} Aix Marseille universit\'e, CNRS, IUSTI UMR 7343, 13453, Marseille, France
}

\authorcopyright{M. Mokhtari, L. Zekri, A. Kaiss, N. Zekri, 2016}
\date{Received May 19, 2016, in final form August 11, 2016}

\begin{document}

\maketitle

\begin{abstract}
The effect of thickness on the width of the percolation threshold in metal-dielectric composite films was examined. The distribution of current intensities through cubic  networks of metal and dielectric components was determined using Kirchhoff's equations. From the tail of current distribution,  the width of the percolation threshold was defined using L\'evy statistics, and determined as a function of the film thickness for a system size 100. In the 2D-3D crossover region, the percolation width decreases as a power-law with a power exponent of $ 0.36\pm 0.01$.
\pacs 05.10.-a, 02.70.-c, 84.37.+q, 51.70.+f, 83.80.Ab, 78.20.-e
\end{abstract}

\section{Introduction}
\vspace{3mm}

Percolation theory is an old model used extensively to describe different second order phase transitions in various fields of disordered systems. Different analytical and numerical methods have been used to determine the percolation threshold and critical exponents, among which there are effective medium theory, field theory, random resistors network, and Monte Carlo simulations \cite{Stauffer, Sahimi, Tuzel, Calabrese, Springett}.

The construction of continuous macroscopic objects having random spreading of particles or links, like polymers, aggregates or piles, or vitrification, as well, have been, so far, the subject of many studies. They were modelled using many different approaches on the basis of the percolation model \cite{Stauffer, Sahimi,Stauffer1979, Yilmaz2002,Yilmaz2002_2,Essam1980}.

The power law behaviours for some physical quantities, such as correlation length and the strength or the weight of the percolating cluster, are investigated near the percolation thresholds (critical points), and the results are interpreted using critical exponents, which depend only on the space dimension D.

Crossover  from 2D to 3D has also been the subject of many studies. For instance, thin magnetic \cite{Jose} and electric \cite{Clerc1980, Sotta2003, Shklovskii,Shklovskii_2} films were investigated. More recently, Zekri et al. \cite{Zekri2011} investigated the effect of thickness on the percolation threshold $p_{\text{c}}$  and the conductivity critical exponent of metal-insulator composites.
The percolation threshold $p_{\text{c}}$ can be defined only for infinite systems where the transition is abrupt. For finite systems, the percolation probability assumes values between $0$ and $1$ for a continuous finite range of the density of ``conducting'' components (see page 72 of \cite{Stauffer}). This defines the width of the percolation threshold.

In the present work, we focus on the width of percolation threshold  dependence on the film thickness. Since the percolation phase transition induces maximum disorder, the width of percolation threshold  is determined from the current distribution by using L\'evy statistics \cite{Levy, Liberman,Liberman_2,Liberman_3}. The novelty of the present work is the use of the width of the percolation threshold to study the percolation in 2D-3D crossover.

\section{Method description}

\begin{figure}[!b]
\centerline{%
\includegraphics[width=10 cm]{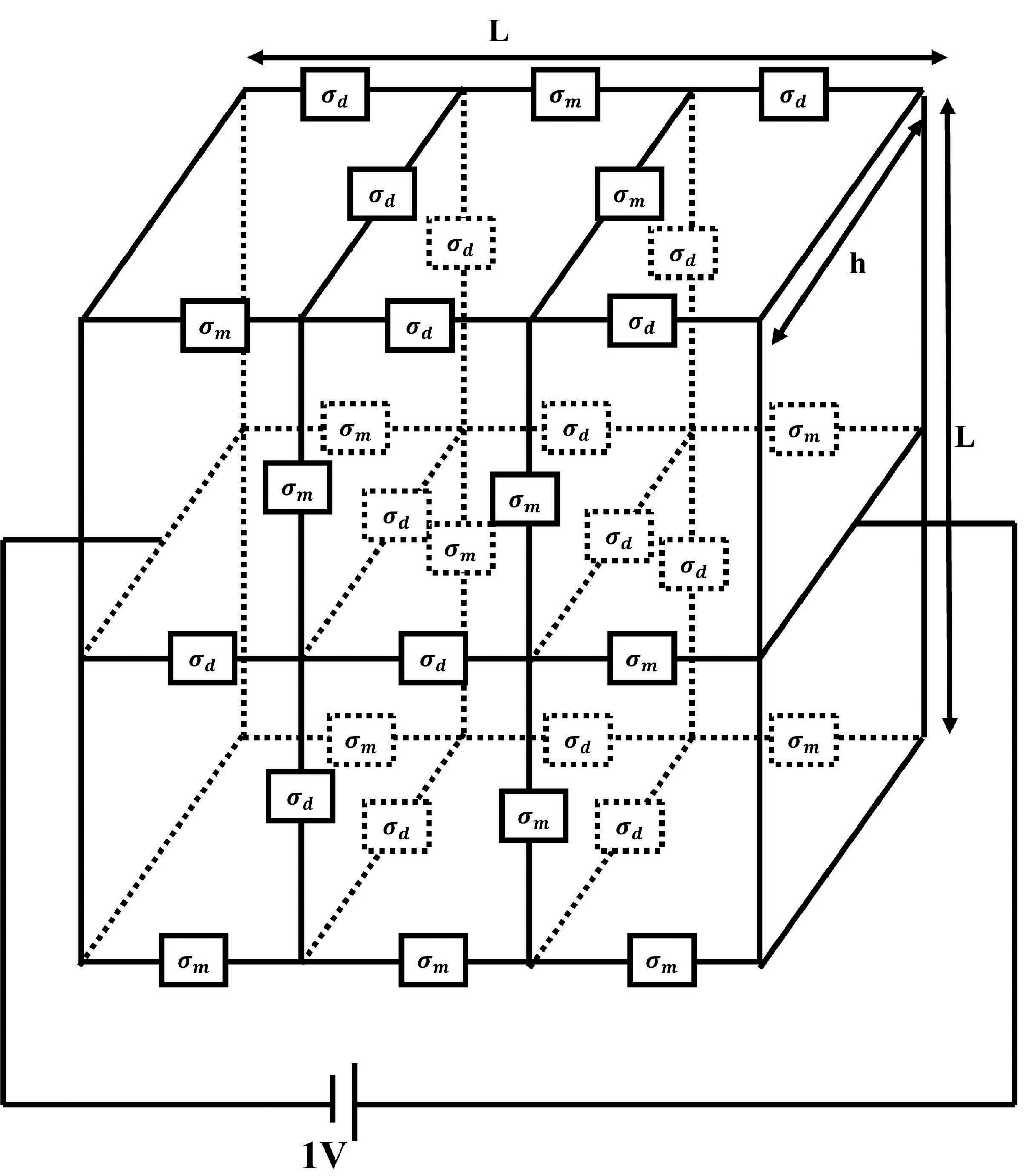}
}
\caption{A network of size $ 3 \times 2 \times 1$  with conductivities distributed randomly between the metals and the insulators.}
\label{Fig1}
\end{figure}

Various calculation methods, such as the real-space renormalization group method (RSRG) \cite{ Sarychev}, exact transformation method like Lobb and Frank method \cite{Lobb}, and transfer matrix method \cite{Derrida1982, Travenec}, are used to numerically define the network conductivity. These methods are limited to 2D or show numerical instability for large systems. We used the exact method (EM) in our previous work \cite{Zekri2004}  based on the numerical resolution of Kirchhoff's equations in the network, leading to the effective conductance determination (admittance or impedance ) and the current or the local field distribution in each component (or network link) as well.
Here we consider a random network of resistors, of $L  \times L\times h$ cubic elementary cells where $h-1$ is the thickness and $L-1$ is the size of the perpendicular plane of the cell (figure~\ref{Fig1}). The network is randomly filled with metallic and insulating components, having conductances of $\sigma_{\textrm{m}} = 1$~\textohm$^{-1}$ and $\sigma_{\text{d}}=10^{-15}$~\textohm$^{-1}$ with filling densities of  $p$ and $1-p$, respectively.

The d.c. voltage in each node and the current in each link are exactly calculated by solving Kirchhoff's equations at each node  $ i,j$ in figure~\ref{Fig1}, for resistor networks \cite{Zekri2004}. The first index represents the coordinate index in the plane and the second one is the plane position in the third direction. The current conservation in each node yields
\begin{equation}
\sum_{k,l} = (V_{k,l}-V_{i,j}) \sigma_{k,l}^{i,j} =0,
\label{1}
\end{equation}
where the sum over $(k,l)$ spans nearest neighbors of the nodes with coordinates $(i,j)$ and conductances~$\sigma_{k,l}^{i,j}$. For the nodes connected to the edges, the potentials $V_{i,j}$ take values $0$ or $1$ depending on the edge side. We have $N$ equations $[N= L \times (L-1) \times  h]$ of $N$ unknown potentials. Equation (\ref{1}) can be written in a matrix form
\begin{equation}
\tilde{\Gamma} \vec{V}= \left[
\begin{array}{ccccc}
P_{11}&P_{12}&0&\ldots&0 \\
P_{21}&P_{22}&P_{23}&\ldots&0 \\
\ldots&\ldots&\ldots&\ldots&\ldots\\
\ldots&\ldots&P_{N-1,N-2}&P_{N-1,N-1}&P_{N-1,N}\\
\ldots&\ldots&0&P_{N,N-1}&P_{N,N}
\end{array}
\right]
\left[
\begin{array}{c}
V_1\\ V_2\\\ldots\\V_{N-1}\\V_N

\end{array}
\right]
=\left[
\begin{array}{c}
S_1\\ S_2\\\ldots\\S_{N-1}\\S_N

\end{array}
\right].
\label{2}
\end{equation}

In this matrix equation, the matrix $\tilde{\Gamma}$ is tri-diagonal symmetrical with $(N - 1)\times (N-1)$ block elements $P_{ij}$ being themselves matrices of dimension $N\times N$. These matrices are composed of combinations of the branch conductivities corresponding to currents directions either within the face $i$ (diagonal block elements) or coming to this face from its neighboring plane faces (off-diagonal block elements). Therefore, at most three of the matrix elements  $P_{ij}$ do not vanish for each face  $ i$. The diagonal matrix is tri-diagonal as it involves the currents within the lines and inter-lines in the same face. The off-diagonal matrix is diagonal. The $N-1$ elements $V_i$ are vectors of size $N$ representing the $N$ unknown potentials. The elements $S_i$ are vectors of the same size as $V_i$. Their elements vanish except the $L$ first ones which correspond to the edge at 1~V. The set of equations (\ref{2}) are solved by the substitution method described in detail in \cite{Zekri2004}.

In this work, in order to further understand the behaviour of the increment between 2D and 3D, we investigate the distribution of the current intensities $P(I_n)$ for different network thicknesses $h$, to evaluate the width of the percolation thresholds $p_{\text{c}}$. This distribution was interpreted on the basis of the L\'evy distribution. The calculations are performed for various thicknesses and many layer sizes. The intensities of currents are averaged by considering 100 samples which is sufficient to reach the desired accuracy \cite{Zekri2004}. Note here that for the networks used here, the percolation threshold for infinite systems is $p_{\text{c}}=0.5$ for 2D square networks ($h=1$), and $ p_{\text{c}}=0.2492$ for 3D cubic network ($h=L=\infty$) \cite{Stauffer}.

\section{Results and discussions}

\begin{figure}[!t]
\centerline{%
\includegraphics[width=13.5cm]{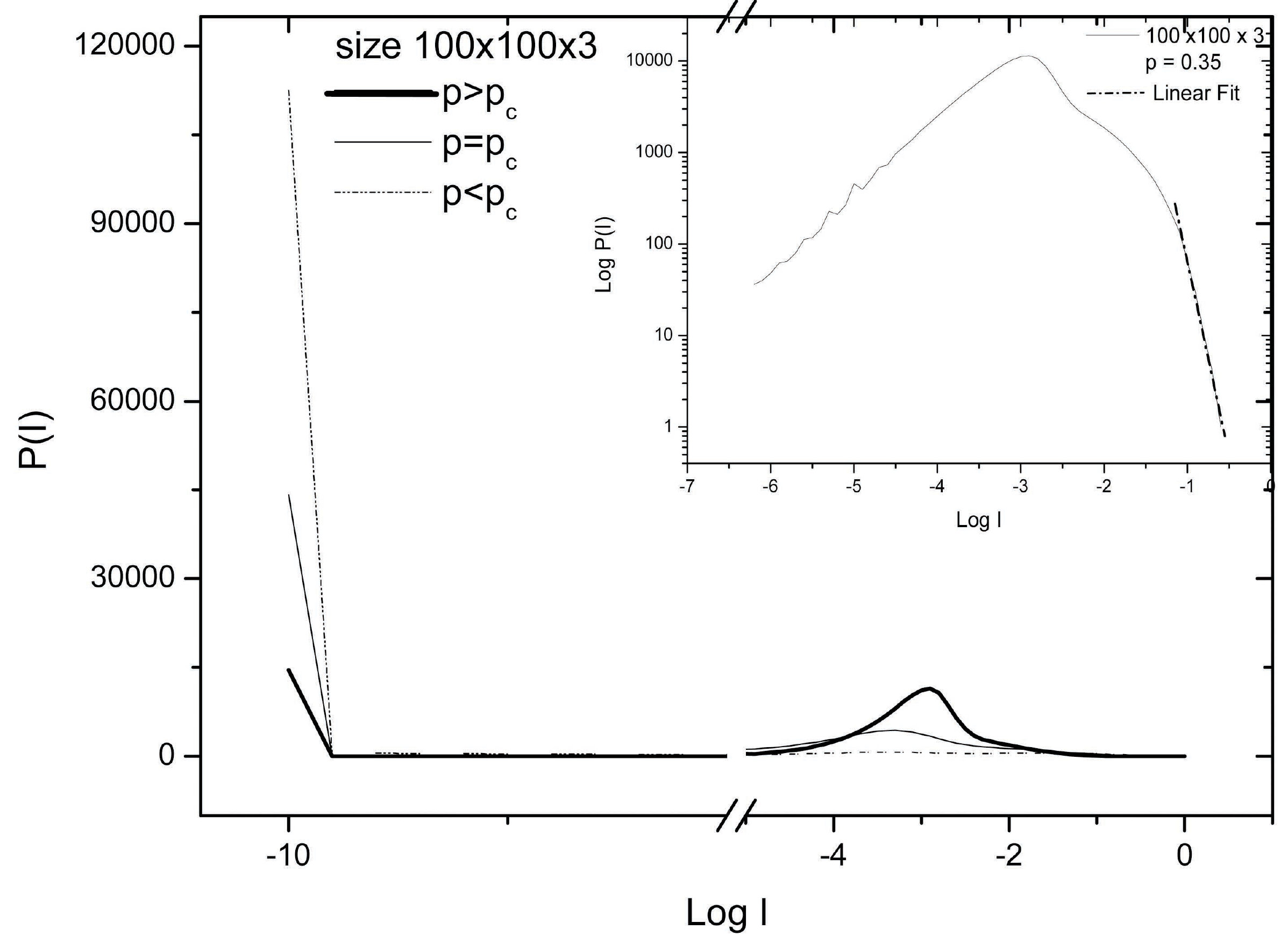}
}
\caption{Distribution of the logarithm of the current for the size  $100 \times 100 \times 3$ and for $p=0.32$, $p_{\text{c}}=0.34$, and $p=0.35$. The inset is a zoom in the region of the log-normal branch.}
\label{Fig2}
\end{figure}

The current distribution is shown in figure~\ref{Fig2} for a system size $100\times 100\times 3$ with values of the metal density below, at and above the percolation threshold. Two main branches appear as the percolation threshold is reached: one seems to be log-normal (Galton's distribution) for large strengths of the current (see the inset of figure~\ref{Fig2}), and the other one is power-law decreasing for very small strengths of the current. The log-normal branch corresponds to the backbone of the metallic largest cluster whereas the power-law decreasing branch corresponds to the insulating cluster.
Below $p_{\text{c}}$, the log-normal branch obviously disappears as there are only small metallic clusters. Above this threshold, the power-law decreasing branch tends to disappear because most of the metallic links belong to the largest cluster. This behaviour is similar to that observed by Zekri et al. \cite{Clerc2005} for the distributions of critical links. Thus, near the percolation threshold, the largest cluster is composed mainly of critical links. These links are responsible for the percolation transition.

The above discussed behaviour of the current distribution is thus related to the percolation phase transition. At the threshold transition, fluctuations (disorder) are larger. In the present case, current fluctuations should be large at the percolation threshold. The behaviour of current fluctuations can be easily determined from L\'evy statistics \cite{Levy, Bouchaud}. Distribution $P(I)$ asymptotically decays as a power law $I^{-(1+\mu)}$, the exponent $\mu$ being positive (see the power-law fit in the inset of figure~\ref{Fig2}). The exponent $\mu$ is then extracted from the linear fit of the log-log plot of $P(I)$ for large values of $I$.

For large (diverging) current fluctuations, the exponent $\mu$ is in the range [0--2]  and the second moment $\langle I^2\rangle$ diverges (the first moment $\langle I\rangle$ diverges for $\mu<1$). For larger exponents ($\mu>2$), the fluctuations converge and the distribution becomes stable \cite{Bouchaud}. In this case, the system is either in its metallic or dielectric phase. Thus, the percolation threshold corresponds to an exponent $\mu$ smaller than~$2$.

\begin{figure}[!t]
\centerline{%
\includegraphics[width=12.6cm]{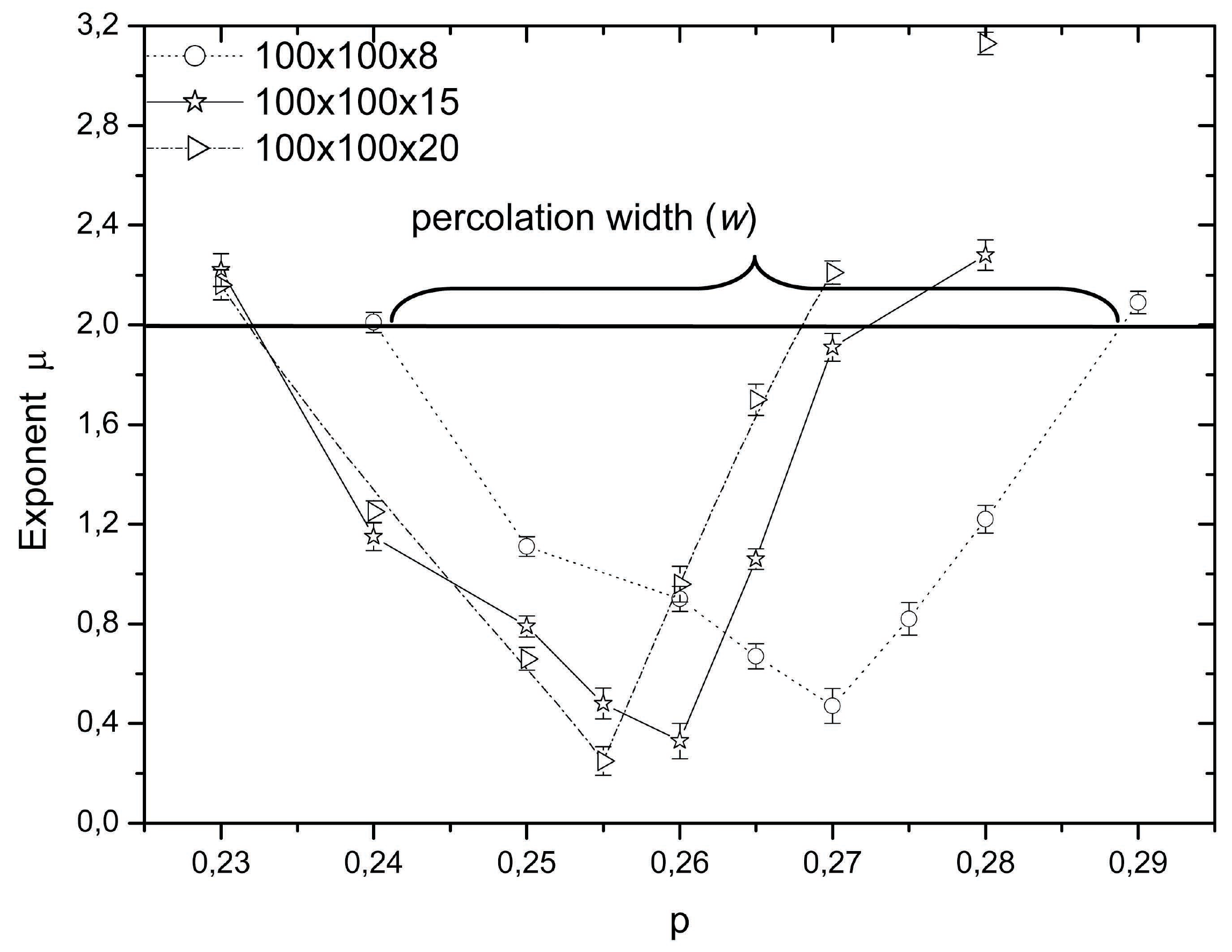}
}
\caption{The exponent $\mu$ versus the concentration $p$  for size $100$ and various thicknesses.}
\label{Fig3}
\end{figure}

 Here, the minimum of the exponent $\mu$ corresponds to the maximum of the current fluctuations (maximum disorder), and thus we define it as the percolation threshold for infinite systems \cite{Stauffer, Zekri2012}. For infinite systems, at this threshold density, only the exponent $\mu$ is smaller or equal to $2$ . For finite systems, fluctuations are large ($\mu \leqslant 2$) for a finite range of densities around $p_{\text{c}}$. This range shows a continuous band of filling densities with diverging current fluctuations, and defines the width of percolation threshold  $w$ for a finite system corresponding to the intersection with the line $\mu=2$  in figure~\ref{Fig3}. Indeed, for a finite system, there is a finite percolation (non-percolation) probability even below (above)  $p_{\text{c}}$ (see page 72 of \cite{Stauffer}).
The threshold values obtained from the minimum of $\mu$ are compared in table~\ref{tbl-smp1} to those obtained in \cite{Zekri2011} for different thicknesses and for $L=100$. These values are comparable within statistical errors. Therefore, as expected above, the percolation threshold for infinite systems corresponds to the filling density with minimum value of the exponent $\mu$ (this density is independent of the size). The use of L\'evy statistics allows one to deduce the percolation threshold for infinite systems by using finite systems.

\begin{figure}[!b]
\centerline{%
\includegraphics[width=12.6cm]{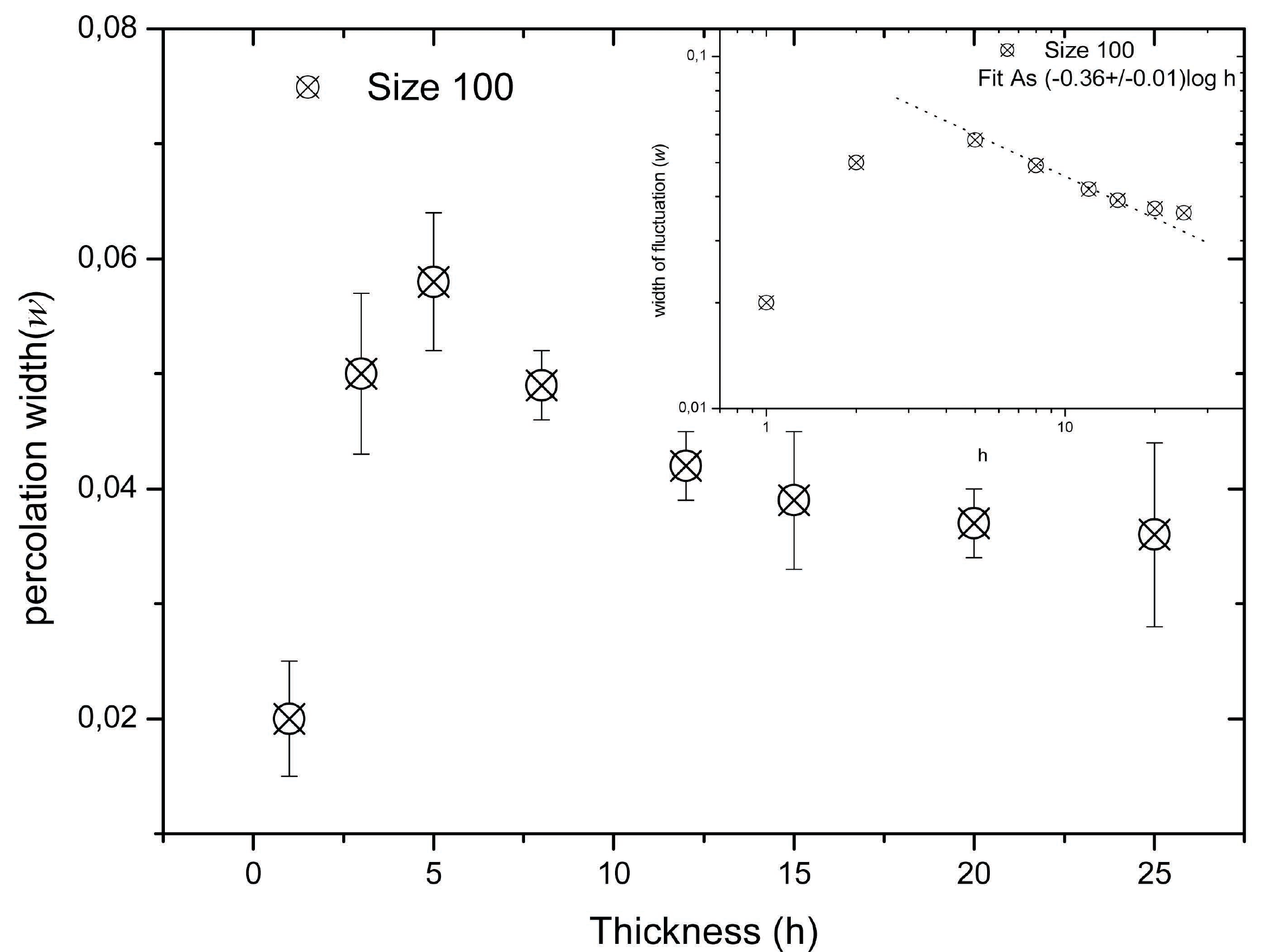}
}
\caption{Percolation width $w$ versus the thickness $h$ for a 3D network of size $L=100$. The inset figure shows the log-log plot of the data.}
\label{Fig4}
\end{figure}

\begin{table}[htp]
\caption{The threshold values as function of the thickness $h$ for the size $L=100$.}

\label{tbl-smp1}

\begin{center}

\begin{tabular}{|c|c|c||c|}
\hline\hline
Thickness $h$ & $p_{\text{c}}$ (present method) & Uncertainty & $p_{\text{c}}$ (method of \cite{Zekri2011}) \\
\hline\hline

3\hphantom{00} & \hphantom{0}0.340 & \hphantom{0}0.005 & \hphantom{0}0.337 \\
4\hphantom{00} & \hphantom{0}0.305& \hphantom{0}0.005 & \hphantom{0}0.305  \\
5\hphantom{00} & \hphantom{0}0.290& \hphantom{0}0.005  & \hphantom{0}0.290\\
8\hphantom{00} & \hphantom{0}0.270 & \hphantom{0}0.005 & \hphantom{0}0.268 \\
10\hphantom{00} & \hphantom{0}0.265& \hphantom{0}0.005 & \hphantom{0}0.262  \\
12\hphantom{00} & \hphantom{0}0.260& \hphantom{0}0.005 & \hphantom{0}0.258 \\
15\hphantom{00} & \hphantom{0} 0.260& \hphantom{0}0.005 & \hphantom{0}0.254\\
20\hphantom{00} & \hphantom{0}0.255 & \hphantom{0}0.005 & \hphantom{0}0.251 \\
25\hphantom{00} & \hphantom{0}0.255 & \hphantom{0}0.005 & \hphantom{0}0.250 \\
\hline\hline

\end{tabular}

\end{center}
\end{table}

 The thickness dependence of the width of percolation threshold is shown in figure~\ref{Fig4} for size $L=100$. In this figure we distinguish three different regions:  a region quasi-2D or small thicknesses $ h < 5$ (where the percolation width $w$ increases), a crossover region, and a saturation where $w$ becomes constant. In the crossover region $(5 \leqslant h \leqslant 10)$, the width of percolation  $w$ is a power-law decreasing function of the $h$ (see the power-law fit in the inset of figure~\ref{Fig4}):
\begin{equation}
w \propto h^{-x}.
\label{moneq}
\end{equation}

	Here, the exponent $x$ is $0.36\pm 0.01$.Within the statistical errors, this value is compatible with $1/\nu_3 -1/\nu_2=0.38$ , where $\nu_2=1.33$  \cite{Stauffer}  and $\nu_3=0.88$ \cite{Sahimi} are 2D and 3D correlation length exponents. The power-law behaviour of the width of percolation threshold $w$ appears similar to that of $p_{\text{c}}(h)$--$p_{\text{c}}(\text{3D})$ with a very close exponent (see equation~(7) of \cite{Zekri2011}).

The non-monotonous behaviour, observed in figure~\ref{Fig4} as $h$ increases, can be explained by the competition between the width $w$ which increases and $p_{\text{c}}$ which decreases. As the thickness increases, there are more percolating paths, so that the width increases. For a size $100$, the width for 3D systems ($0.039$) is greater than that for 2D ones ($0.02$). For very small thicknesses ($h<5$), the increase of $w$ is independent of the decrease of $p_{\text{c}}$. Above $h>5$, $p_{\text{c}}$ significantly decreases and thus the width becomes constrained by $p_{\text{c}}$. The quantity $p_{\text{c}}\times w$ seems to be constant for $h<5$ and decreases above this thickness (not shown here). A further work is under preparation to clarify this behaviour near the crossover region.

\section{Conclusion}
\vspace{3mm}

In this work, the width of  percolation threshold of a random metal-dielectric composite network was interpreted from the behaviour of the current distribution using L\'evy statistics. The percolation threshold~$ p_{\text{c}}$ corresponds to the minimum of the exponent $\mu$. The width of the percolation threshold $w$ was examined as a function of the thickness $h$. A power-law decrease of $w$ is found in the crossover region with an exponent similar to that obtained for $p_{\text{c}}(h)$--$p_{\text{c}}(\text{3D})$ in \cite{Zekri2011}. This induces a relation between the width $w$ and $p_{\text{c}}(h)$. The saturation region corresponds to the 3D behaviour. As discussed above, the width $w$ should vanish for infinite systems. The scaling of this width for different thicknesses will be the subject of a forthcoming work.

\ukrainianpart

\title{Вплив товщини плівки на ширину перколяційного порогу в композитах метал-діелектрик}

\author{M. Мохтарі\refaddr{label1,label2}, Л. Зекрі\refaddr{label1}, А. Кесс\refaddr{label3}, Н. Зекрі\refaddr{label1} }
\addresses{
\addr{label1} Університет науки і технологій ім. Мохамеда Будіафа м. Оран, USTO-MB, LEPM, \\BP 1505, 31000 Оран, Алжир
\addr{label2} Університетський центр  Тіссемсілту, BP 182, Тіссемсілт, Алжир
\addr{label3} Марсельський університет, CNRS, IUSTI UMR 7343, 13453, Марсель, Франція
}

\makeukrtitle

\begin{abstract}
Досліджено вплив товщини плівки на ширину перколяційного порогу у плівках метал-діелектричних композитів.
Розподіл інтенсивностей струму через кубічні мережі компонентів металу і діелектрика визначався, використовуючи рівняння Кіргофа.
Ширина перколяційного порогу визначалася з хвоста розподілу струму, використовуючи статистику Леві, і була визначена як функція товщини
плівки для розміру системи 100. В кросоверній області 2D-3D ширина перколяції зменшується за степеневим законом з показником  $ 0.36\pm 0.01$.
\keywords розподіл струму, перколяція, композитні матеріали, резисторна мережа
\end{abstract}

\end{document}